\documentclass[9pt,sigconf]{acmart}
\usepackage{siunitx}
\usepackage{xspace}
\usepackage{tikz}
\usepackage{mdframed}

\usepackage{balance}

\usepackage{url}
\usepackage{breakurl}

\usepackage{atbegshi}

\usetikzlibrary{arrows.meta,automata,positioning,shapes.misc,shadows}

\def\system{{\sc PfIP}\xspace}
\def\openmac{ESP32-Open-MAC\xspace}
\def\platform{ESP32-C3\xspace}

\newcommand{\highlightcircled}[2]{%
  \tikz[baseline=(char.base)]\node[circle, fill=#2, inner sep=0.5pt, minimum size=1mm, text=white, outer sep=0pt](char){#1};%
}

\definecolor{respectBlue}{HTML}{0071BC}
\definecolor{respectGreen}{HTML}{009B55}
\definecolor{respectRed}{HTML}{ED1B23}
\newcommand{\sleepstate}{\highlightcircled{1}{respectBlue}\xspace}
\newcommand{\idlestate}{\highlightcircled{2}{respectGreen}\xspace}
\newcommand{\txstate}{\highlightcircled{3}{respectRed}\xspace}

\AtBeginDocument{%
  }

\copyrightyear{2025}
\acmYear{2025}
\setcopyright{cc}
\setcctype{by-sa}
\acmConference[ENSsys '25]{The 13th International Workshop on Energy Harvesting and Energy-Neutral Sensing Systems }{May 6--9, 2025}{Irvine, CA, USA}
\acmBooktitle{The 13th International Workshop on Energy Harvesting and Energy-Neutral Sensing Systems (ENSsys '25), May 6--9, 2025, Irvine, CA, USA}
\acmDOI{10.1145/3722572.3727926}
\acmISBN{/2025/05}
\newcommand\AuthorsVersionMarker{
  \AtBeginShipoutNext{\AtBeginShipoutUpperLeftForeground{
\begin{tikzpicture}[remember picture,overlay]
\centering
\node [fill=yellow!20, text width=\textwidth, rounded corners, anchor=north, yshift=-1.8em] at (current page.north) {\sffamily\footnotesize{
\hspace{-0em}{\textbf{Appears in:}\hfill{} Proceedings of the 13th International Workshop on Energy Harvesting and Energy-Neutral Sensing Systems (ENSsys '25)\\ \hfill{} May 6--9, 2025, Irvine, CA, USA}}};
\end{tikzpicture}
}}}%

\begin{document}

\title{Reverse Engineering the ESP32-C3 Wi-Fi Drivers for Static Worst-Case Analysis of Intermittently-Powered Systems}

\author{Ishwar Mudraje}
\affiliation{%
  \institution{Saarland Informatics Campus (SIC)}
  \city{Saarbr{\"u}cken}
  \country{Germany}
}

\author{Kai Vogelgesang}
\affiliation{%
\institution{Saarland Informatics Campus (SIC)}
\city{Saarbr{\"u}cken}
\country{Germany}
}

\author{Jasper Devreker}
\affiliation{
  \institution{}
  \city{Ghent}
  \country{Belgium}
}

\author{Luis Gerhorst}
\affiliation{%
\institution{Friedrich-Alexander-Universit{\"a}t}
\city{Erlangen-N{\"u}rnberg}
\country{Germany}
}

\author{Phillip Raffeck}
\affiliation{%
\institution{Friedrich-Alexander-Universit{\"a}t}
\city{Erlangen-N{\"u}rnberg}
\country{Germany}
}

\author{Peter W{\"a}gemann}
\affiliation{%
\institution{Friedrich-Alexander-Universit{\"a}t}
\city{Erlangen-N{\"u}rnberg}
\country{Germany}
}

\author{Thorsten Herfet}
\affiliation{%
\institution{Saarland Informatics Campus (SIC)}
\city{Saarbr{\"u}cken}
\country{Germany}
}

\renewcommand{\shortauthors}{Mudraje et al.}

\begin{abstract}
The Internet of Batteryless Things revolutionizes sustainable communication as it operates on harvested energy. This harvested energy is dependent on unpredictable environmental conditions; therefore, device operations, including those of its networking stack, must be resilient to power failures. Reactive intermittent computing provides an approach for solving this by notifications of impending power failures, which is implemented by monitoring the harvested energy buffered in a capacitor. However, to use this power-failure notification and guarantee forward progress, systems must break down tasks into atomic transactions that can be predictably finished before the energy runs out. Thus, static program-code analysis must determine the worst-case energy consumption (WCEC) of all transactions. In Wi-Fi--capable devices, drivers are often closed-source, which avoids the determination of WCEC bounds for transactions since static analysis requires all code along with its semantics.

In this work, we integrate an energy-aware networking stack with reverse-engineered Wi-Fi drivers to enable full-stack WCEC analysis for physical transmission and reception of packets. Further, we extended a static worst-case analysis tool with a resource-consumption model of our Wi-Fi driver. Our evaluation with the RISC-V--based ESP32-C3 platform gives worst-case bounds with our static analysis approach for the transactions of the full communication stack, therefore showing that Wi-Fi--based reactive intermittent computing is feasible.

\end{abstract}




\maketitle
\AuthorsVersionMarker
\section{Introduction}
\paragraph{Batteryless Systems}
Batteryless devices~\cite{ahmed:2024:cacm} rely on harvested energy.
Being batteryless not only results in independence from battery requirements but also makes novel energy-self--sufficient applications possible:
Batteryless medical implants~\cite{kim:2024:adma} reduce the need for surgical replacements due to degradation in battery performance, putting them into the scope of safety- and security-critical systems~\cite{mottola:2024:eurosec}.
Naturally, the unpredictable nature of harvesting energy means that a started task can face a power failure before reaching a checkpoint stored in non-volatile memory.
Power failures cause two major problems:
First, failures can leave the system in an inconsistent state.
Second, with failures during energy-intensive operations with transactional semantics~(e.g., sending packets), systems might repeatedly face failures without making progress.

\paragraph{Worst-Case Analysis}
To address inconsistency and starvation, static worst-case program analysis can guarantee power-failure--free execution in intermittent systems.
Specifically, static \emph{worst-case energy consumption~(WCEC)} analysis~\cite{choi:2022:rtas,raffeck:2024:lctes,waegemann:2018:ecrts,yarahmadi:2020:samos} determines a safe upper bound on operations' energy demand.
These analysis techniques originally stem from determining the \emph{worst-case execution time~(WCET)} to guarantee timeliness in real-time systems~\cite{wilhelm:2008:tecs}.
With WCEC estimates and the available energy in the storage~(i.e., capacitor) during runtime, operations execute without the risk of power failures.
Consequently, consistency and forward progress are guaranteed under consideration of the currently available energy.

\paragraph{Available Code \& Semantics}
A necessary requirement for such static resource-consumption analyses is the availability of all code along with its resource-consumption semantics~\cite{abella:2015:sies,schoeberl:2009:eurasip}.
This knowledge is particularly important for power-consuming operations, such as radio transceivers, since inaccuracies in the static behavior modeling can result in unsound or impractically pessimistic analysis estimates.
Unfortunately, for our target platform, the RISC-V--based \platform system-on-chip~(SoC), the Wi-Fi drivers are closed-source, and the Wi-Fi peripheral's behavior is undocumented.

\paragraph{Contributions}
The \openmac project~\cite{devreker:2023:unveiling} provides re\-verse-engineered Wi-Fi drivers for an Xtensa-based SoC, which we leveraged for the RISC-V--based ESP32-C3.
Our work presents open-source Wi-Fi drivers for the~\platform alongside their static worst-case analyzability.
In summary, our contributions are threefold:
\begin{enumerate}
    \item \emph{Case Study on Reverse Engineering}: We present our reverse engineering efforts for the Wi-Fi driver of the~\platform. %
    \item \emph{Energy-Consumption \& Execution-Time Model for Worst-Case Analysis}: Our static analysis comes with a fine-grained en\-ergy-consumption model. %
    \item \emph{Open-Source Implementation \& Evaluation}: %
    Using the reverse-engineered knowledge, we develop an open-source Wi-Fi driver for the \platform and present evaluation results that validate the driver's property of being statically analyzable.
\end{enumerate}

\section{Background}\label{sec:background}
This section gives background on network stacks and mechanisms for processor and Wi-Fi module communication.
Finally, we outline worst-case resource-consumption analyses of program code. %

\subsection{Network Stack}
The network stack on embedded devices typically consists of a lightweight TCP/IP implementation and a driver that interacts with the physical network interface (e.g., Wi-Fi peripheral).
At the highest level of the stack is the application that provides data to be transmitted, while at the lowest level is the physical layer (PHY), which sends out the data onto the physical medium.
On top of the physical layer is the medium access control~(MAC) that coordinates transmission between multiple devices.
In the case of Wi-Fi, the carrier sense multiple access/collision avoidance~(CSMA/CA) mechanism is used to coordinate transmissions.
Each layer of the network stack encapsulates data from the upper layer and passes it to the layer below.
On intermittent devices, power may be lost at any point before packet transmission is completed, leading to a loss of progress up to that point.

In the case of Wi-Fi, PHY-layer operations are implemented in hardware due to strict timing requirements.
Some Wi-Fi peripherals implement MAC operations in software (SoftMAC), while others implement both MAC and PHY operations in hardware (FullMAC).

\subsection{Wi-Fi Hardware}
High-speed communication between the Wi-Fi peripheral and the processor within an SoC is typically carried out using a combination of memory-mapped I/O~(MMIO) and direct memory access~(DMA)~\cite{schulz:2018:wifi}.
The Wi-Fi peripheral is configured through MMIO, while data transfer for transmission and reception is performed using DMA.
DMA allows the CPU to remain free for other tasks instead of handling the transfer.
Espressif devices such as~\platform follow this architecture.
The peripheral communicates events such as the successful transmission or reception of a packet through interrupts.
On other platforms, external Wi-Fi chipsets may use the SPI interface to communicate with the host controller.

\subsection{Static Worst-Case Resource Analysis}
Static worst-case resource-consumption analysis of program code comprises two major pillars:
(1)~a hardware-agnostic program path analysis and (2)~a hardware-dependent cost modeling capturing the actual resource demand~(i.e., execution time or energy demand) for the actual target platform.
The main idea is to formulate a mathematical problem formulation, usually an integer linear program~(ILP):
The ILP's constraints originate from the path analysis.
Abstract-interpretation techniques are further useful for refining the path constraints in a context-sensitive manner.
The ILP's objective function states the maximization of the actual costs with the given path constraints.
Eventually, the ILP's solution yields a mathematically proven worst-case bound.
With regard to energy consumption, the WCEC bound is useful to guarantee the execution without power failures during the runtime of batteryless devices.

\section{Problem Statement}\label{sec:problem-statement}
The problems that we address in this work are twofold:
Section~\ref{sec:problem:driver} discusses the problem of closed-source drivers,
and Section~\ref{sec:problem:model} explains the requirement of resource models.

\subsection{Problem of Closed-Source Wi-Fi Drivers}
\label{sec:problem:driver}

In intermittently-powered devices, static worst-case analysis techniques can be crucial to gain knowledge on the worst-case energy consumption of operations~\cite{choi:2022:rtas,raffeck:2024:lctes,yarahmadi:2020:samos}.
During runtime, the system's scheduler checks the available energy~(comparable to just-in-time checkpointing) and reactively dispatches tasks only if sufficient energy is available to guarantee freedom from power failures.
The WCEC-aware approaches give both memory-consistency and forward-progress guarantees.
The trade-off we exploit favors expensive design-time analyses over complicated runtime checkpointing mechanisms.
Once deployed, systems achieve provable resilience against power failures.
However, these design-time analyses require the code to be available.
Our target platform, the~\platform, is a RISC-V--based, widely used SoC from Espressif, which has closed-source Wi-Fi drivers.
Although less power-intensive networking technology exists~(e.g., BLE), we are interested in the Wi-Fi stack being of high relevance:
In 2024, an estimated 31\,\% of IoT devices used Wi-Fi for connectivity~\cite{sinha:2024:stateiot}. %
Further, the recent article on \emph{The Internet of Batteryless Things}~\cite{ahmed:2024:cacm} calls for action and states that popular network protocols, including Wi-Fi, should be adapted for intermittent operation.
Unfortunately, for our target, the generic Wi-Fi drivers were not available as analyzable source code, which is a common problem in embedded SoCs.
Of main interest for our analysis are the MAC and PHY layers.

\emph{Our Approach in a Nutshell:} To solve the problem of the closed source, we exploit several reverse-engineering techniques on Espressif's closed-source Wi-Fi driver alongside utilizing existing knowledge~\cite{devreker:2023:unveiling}.
Fortunately, Espressif legally allows reverse engineering this driver.
Based on the reverse-engineering results, we contribute an open-source driver for the ESP32-C3.

\subsection{Missing Resource-Consumption Model}
\label{sec:problem:model}
Although the developed open-source driver gives us the freedom to perform static analysis, a major problem is connecting memory-mapped IO operations to semantics with regard to the power demand.
When considering the following MMIO operation:\newline
\texttt{*((uint32\_t*)0x60033ca0) \&= 0xff00efff},\newline
the static analysis is not aware of its semantics. %
On the~\platform, this operation switches the transceiver from a sleep state to standby mode~(see also Figure~\ref{fig:dev-graph}).
Hence, our static resource analysis needs to include the notion that the power drawn increases after this operation.
Similarly, specific instructions also require a longer execution time than a normal RAM access, which in turn leads to higher energy demand~(i.e., power over time).
Consequently, a comprehensive model needs to account for power, energy, and time.

\emph{Our Approach in a Nutshell:} For our target, we present a resource-consumption model that considers transitions in the hardware, which change the SoC's power demand.
Our analysis makes use of this model and tracks program paths in a context-sensitive way.

\section{Approach}\label{sec:approach}
We use a combination of commonly available static and dynamic reverse-engineering analysis methods for identifying the registers and, correspondingly, their contents.

\subsection{Reverse Engineering}
\paragraph*{Static Analysis} Through decompilation, the program flow and, accordingly, the registers used for transmission and reception can be identified.
Functions within the binaries are named based on their purpose, which leads to informative decompilation results.
Additionally, the undocumented interrupt configuration used by the Wi-Fi peripheral can be identified.
However, decompilation alone is insufficient to identify control and data flows.
Several of the registers are configured based on data that is internal to the driver, whose structure is not documented.
Moreover, the peripheral also has several slots for transmission, which are dynamically selected based on the network configuration.
Therefore, we use QEMU to dynamically determine the register contents as well as the slots.

\paragraph*{Dynamic Analysis} The QEMU version in~\cite{gamboa:2023:qemu}, a fork of Espressif's QEMU, supports Wi-Fi for \platform by emulating an open Wi-Fi access point~(AP). %
In particular, it only implements registers required for triggering a transmission and reading received packets through DMA as well as the status register, which encodes the reasons for an interrupt occurrence.
We further modify this fork to log all register accesses and their contents within the Wi-Fi peripheral's address range.
While the QEMU fork provides insights into accesses of the peripheral's registers and their contents used during the emulation, it does not fully emulate the hardware. 
Hardware initialization, for example, is not emulated.
Moreover, it is not always possible to deduce the purpose of the register operation.

\paragraph*{Binary Hooking for Injecting Tracing Code}
The modules of Espressif's Wi-Fi driver are provided as archives (\texttt{.a}) of object (\texttt{.o}) files.
These object files have undefined symbols referring to functions defined in other modules, such as the binary blobs of the ESP32-C3 ROMs.
Their definition is resolved during the final linking of the executable.
To trace the execution, we hook into these symbols by replacing the symbol names and exploiting the original symbol names in our custom firmware to identify the control flow.

These reverse-engineering methods, in combination with the existing infrastructure in~\openmac, yield the necessary information for constructing the Wi-Fi driver for the \platform.

\subsection{The Open-Source ESP32-C3 Driver}

\begin{figure}
    \centering
    \begin{tikzpicture}[task/.style={rectangle, drop shadow, draw=black, fill=respectBlue!20, text opacity=1, fill opacity=1.0, very thick, minimum size=5mm, font=\sffamily}]
    \tikzstyle{arrow} = [thin,->,>=stealth]
    \usetikzlibrary{positioning}

    \node[task] (app) {Application};
    \node[task] (ip) [right=1.25cm of app] {IP Stack};
    \node[task] (mac) [below=1.0cm of ip] {MAC layer};
    \node[task] (wifi) [below=1.0cm of app] {Wi-Fi driver};
    \node[task] (hw) [left=1.25cm of wifi, align=center] {Wi-Fi\\peripheral};

    \draw [arrow] (app.south) -- (wifi.north) node[midway, above, rotate=90] () {Init};

    \draw [arrow] ([yshift=2mm]app.east) -- ([yshift=2mm]ip.west) node[midway,above] () {API};
    \draw [arrow] ([yshift=-2mm]ip.west) -- ([yshift=-2mm]app.east) node[midway,below] () {Rx CB};

    \draw [arrow] ([xshift=-2mm]ip.south) -- ([xshift=-2mm]mac.north) node[midway, above, rotate=90] () {API};
    \draw [arrow] ([xshift=2mm]mac.north) -- ([xshift=2mm]ip.south) node[midway, below, rotate=90] () {Rx CB};

    \draw [arrow] ([yshift=2mm]wifi.east) -- ([yshift=2mm]mac.west) node[midway,above] () {Queue};
    \draw [arrow] ([yshift=-2mm]mac.west) -- ([yshift=-2mm]wifi.east) node[midway,below] () {Queue};

    \draw [arrow] ([yshift=2mm]wifi.west) -- ([yshift=2mm]hw.east) node[midway,above,align=center] () {MMIO/\\DMA};
    \draw [arrow] ([yshift=-2mm]hw.east) -- ([yshift=-2mm]wifi.west) node[midway, below] () {Interrupt};
    
\end{tikzpicture}
    \caption{Architecture of the firmware.}
    \label{fig:arch}
    \Description[architecture]{architecture of firmware}
\end{figure}
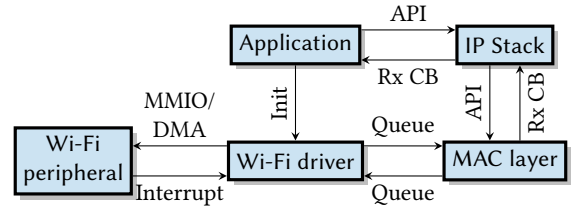

Once the registers and transmit/receive mechanisms are identified, we use the architecture shown in Figure~\ref{fig:arch} to develop the driver.

\paragraph*{MAC Task} The MAC task accepts packets in Ethernet (802.3) format from the IP layer.
It adds the Wi-Fi MAC header and passes it to the driver for transmission.
On the other hand, packets arriving from the driver are passed to the IP layer through a callback~(CB).
The MAC task provides a transmit primitive for the IP stack to queue packets for transmission.
Conversely, a receive callback has to be provided by the IP stack.

\paragraph*{Driver Task} The driver task interacts with the MMIO registers and manages memory allocated to DMA structures required by the peripheral.
All memory is statically allocated to aid WCET/WCEC analysis.
Processing of interrupts is deferred by the interrupt handler to the driver task.
Incoming 802.11 packets are passed to the MAC task for further processing through a queue.
The interrupt handler notifies the driver of the occurrence and reason for the interrupt using a hardware event queue.
The MAC task also writes packets to be transmitted into this queue.
On the other hand, the driver task writes any received packets into the MAC event queue to signal the MAC task of received packets.

\subsection{Building a Resource-Consumption Model}
\label{ssec:approach:model}
For device operations such as transmission and reception, the Wi-Fi peripheral operates independently of the CPU.
To yield sound results, static resource analysis requires a cost model specifying the maximum time and energy demands of operations.
These paragraphs provide a generic overview of the cost model, while Section~\ref{sec:eval_results} discusses concrete values.

\paragraph*{Timing Information}
The necessary timing-related costs comprise both protocol- and hardware-specific aspects.
The application-specific protocol influences the maximum duration of device operations~(e.g., through the maximum packet size).
Additionally, timeouts and operation windows in the protocol allow deriving an upper bound on operations' execution times, such as waiting to receive a packet.
Bounds on the duration of operations also enable bounding the occurrence of interrupts related to the completion of a transmission or an incoming packet.
The second aspect, the hardware-specific behavior, concerns the Wi-Fi peripheral's MMIO interface.
The execution time of certain MMIO operations, such as waiting for a bit to change its value, which signals the completion of a configuration update, depends on the reaction of the hardware.

\begin{figure}
    \centering
    \begin{tikzpicture}
      [shorten >=1pt, node distance=0.5cm, >={Stealth[round]}, initial text=, 
       every state/.style={rectangle, drop shadow, rounded corners, draw=black, very thick, fill=respectGreen!20, align=left, anchor=west, inner sep=5pt, font=\sffamily},
       accepting/.style=accepting by arrow]

  \node[state]  (sleep) {\sleepstate Sleep};
  \node[state]  (standby) [right=of sleep] {\idlestate Standby};
  \node[state]  (tx) [right=of standby] {\txstate Transmitting};

  \path[->] (sleep.north east) edge [bend left] node {} (standby.north west)
            (standby.south west) edge [bend left] node  {} (sleep.south east)
            (standby.north east) edge [bend left] node  {} (tx.north west)
            (tx.south west) edge [bend left] node  {} (standby.south east);
\end{tikzpicture}
    \caption{Simplified device graph, used in the static resource analysis, that holds power states of the Wi-Fi peripheral.}
    \label{fig:dev-graph}
    \Description[device graph]{device graph showing energy states}
\end{figure}
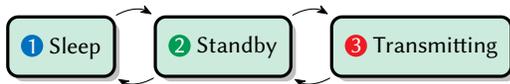

\paragraph*{Energy Information}
The energy-related cost model for the Wi-Fi peripheral consists of all possible device states with distinct power consumption.
A simplified version of the device graph modeling the states and their transitions for the Wi-Fi peripheral is shown in Figure~\ref{fig:dev-graph}.
The depicted graph features three states of operation: \sleepstate Sleep, \idlestate Standby, and \txstate Transmitting.
The Wi-Fi peripheral is powered down and inoperative in the sleep state~(\sleepstate).
In the standby state~(\idlestate), the Wi-Fi peripheral is powered up and may receive packets. 
Lastly, in the transmitting state~(\txstate), the device actively transmits a packet.
No reception occurs in this state, as the peripheral owns the channel through the CSMA/CA mechanism.
For this paper, we assume that the CPU is always on and uses only one run mode, although modeling the ESP32-C3's sleep states, run modes, and clock-frequency changes is generally possible~\cite{dengler:2023:ecrts}.

\subsection{Implementation}
\paragraph*{Reverse Engineering}
The \openmac project lists the necessary MMIO registers to achieve basic transmission and reception on the ESP32.
On the ESP32, the MMIO addresses belong to the~\texttt{0x3ff73000}-\texttt{0x3ff75fff} range, but on the~\platform are mapped to \texttt{0x60033000}-\texttt{0x60035fff}.
The offsets of \platform's MMIO registers from the base address differ from those of the ESP32.
In order to identify the equivalent registers on the~\platform, we write a firmware that performs simple UDP transmission and analyze it using Ghidra's~\cite{nsa:2019:ghidra} decompiler.
In Ghidra, we mark the memory ranges of the peripheral identified with QEMU~\cite{gamboa:2023:qemu} as volatile and identify the functions that access these memory locations. 
Functions with prefixes such as \texttt{hal\_mac\_*} or \texttt{wdev\_mac\_*} were found to handle MAC/PHY layer operations, which are of particular interest for our reverse engineering.

Among the identified functions, we hook into those that are listed as undefined symbols in object files within the \texttt{libpp.a} archive of the proprietary blobs.
The resulting firmware with the hooks is loaded into our QEMU fork.
\begin{table}
\begin{tabular}{|c|p{0.5\linewidth}|c|}
    \hline
    Register & Deduced Purpose & Access\\
    \hline
    \texttt{0x60033000} & MAC address configuration & R/W\\
    \hline
    \texttt{0x60033084} & MS bit enables or disables Rx& R/W\\
    \hline
    \texttt{0x60033088} & Set base DMA linked list address & R/W\\
    \hline
    \texttt{0x60033ca0} & Power up and power down Wi-Fi module & R/W\\
    \hline
    \texttt{0x60033d08} & Configure Tx DMA address and trigger Tx & R/W\\
    \hline
    \texttt{0x60033c3c} & Get interrupt reason & R\\
    \hline
    \texttt{0x60033c40} & Clear interrupt & R/W\\
    \hline
    \texttt{0x60033cac} & Clear Tx slot & R/W\\
    \hline
\end{tabular}
\caption{\platform Wi-Fi peripheral addresses~(excerpt).}
\label{tab:wifi_reg}
\end{table}
An excerpt of the most important registers identified through our reverse engineering is shown in Table~\ref{tab:wifi_reg}, along with the deduced purpose of these registers.

\paragraph*{Interrupt Handling} %
The details of the Wi-Fi interrupt handling are not exposed to developers.
Through decompilation, we find that the Wi-Fi driver uses CPU interrupt number 1, which is marked as reserved in the manual~\cite{esp32c3:trm}, and uses an undocumented register for that interrupt source at address \texttt{0x600c2000}.
With this information, registering a Wi-Fi interrupt handler follows the usual procedure on the \platform.
First, the (undocumented) interrupt source register (\texttt{0x600c2000}) is cleared, and the interrupt is disabled by clearing bit 1 of the (documented) interrupt-enable register~(\texttt{0x600c2104}). 
The interrupt handler is registered using ESP-IDF's \texttt{intr\_handler\_set}.
Finally, the interrupt is enabled by setting bit 1 of \texttt{0x600c2104} and loading the interrupt number to~\texttt{0x600c2000}.

\paragraph*{\platform Wi-Fi Driver} We implement the driver using the architecture shown in Figure~\ref{fig:arch}.
We use an open AP, as the encryption functionality of the Wi-Fi hardware is not fully reverse-engineered.
For initialization and connecting to an AP, we use ESP-IDF's API.
The functions \texttt{esp\_wifi\_init} and \texttt{esp\_wifi\_start} initialize and configure the Wi-Fi, set the internal interrupt handler for Wi-Fi, and initiate an AP connection.
Their successful completion can be monitored through the system notification \texttt{WIFI\_EVENT\_STA\_CONNECTED}.
Once the notification is received, the Wi-Fi task can be killed using the undocumented \texttt{pp\_post} function, which queues instruction codes to the Wi-Fi task started by the driver.
Finally, our driver and MAC tasks are started.
We use~\system~\cite{vogelgesang:2025:pfip} as the IP stack due to its intermittency-aware execution. 

\section{Evaluation}\label{sec:evaluation}
We perform a static resource analysis of our Wi-Fi driver to assess the ability to bound the program flow and resource demand.
Section~\ref{sec:eval_setup} describes the hardware and network setup as well as the task system used for the experiments.
Additionally, we detail the required assumptions for the static resource-consumption analysis using the Platin analysis framework~\cite{maroun:2024:wcet}.
In Section~\ref{sec:eval_results}, we examine and discuss the analysis results.

\subsection{Evaluation Setup}
\label{sec:eval_setup}

\paragraph*{Hardware Setup}
The target platform for evaluation is the ESP32-C3-DevKitM-1 board, featuring an ESP32-C3-MINI-1 module integrating an ESP32-C3 RISC-V processor and a Wi-Fi transceiver.
The board is powered via USB, while an additional voltage regulator provides a stable voltage of \qty{3.3}{V}.
The clock frequency of the processor is fixed at \qty{160}{\mega\hertz}.
The evaluation tasks run on top of the runtime of the ESP IoT Development Framework~(ESP-IDF)~\cite{espressif:esp-idf}, which builds upon the FreeRTOS operating system.

We configure an access point to host an open Wi-Fi network in the \qty{2.4}{GHz} band with static IP allocation for the target communication devices.
802.11 packets sent within the network are monitored using Wireshark~\cite{wireshark}.
The \platform is configured to transmit at \qty{54}{Mbps}.
Wi-Fi features such as frame aggregation are disabled.

\paragraph*{Functionality}
To validate the driver's functionality, we write a firmware that sends UDP packets through~\system's network stack.
\system is provided with the \texttt{mac\_task}'s transmit handle, which in turn queues the packet for physical transmission by the~\texttt{hw\_task}.
The responses from the network are passed back through~\system's receive callback.
We capture the network traffic using Wireshark to monitor our driver's functional correctness.

\paragraph*{Static Resource Analysis}
To analyze the worst-case resource demand of our Wi-Fi driver, we use the Platin analysis framework~\cite{maroun:2024:wcet}.
With the device-model analysis of the WoCA approach~\cite{raffeck:2024:lctes}, Platin supports device-aware resource-consumption analysis for the ESP32-C3.
Platin uses a modified version of the clang compiler that collects user-annotated knowledge, derives metainformation about the program flow, and tracks the relation between source and object level regarding this control-flow information.
Using this metainformation and hardware-specific cost models, Platin formulates the costs of and constraints on the program behavior as an ILP suitable for common mathematical solvers, such as lp\_solve~\cite{lpsolve} or Gurobi~\cite{gurobi}, to obtain the maximum resource demand.
For the purpose of this paper, we annotate the current consumption for each state in the device's energy model based on the values provided by the manufacturer in the datasheet~\cite{esp32c3:2022:ds}.
For the processor, we rely on the model for the ESP32-C3 already present in Platin.

The focus of our evaluation lies on the driver functions, independent of the runtime and operating system.
While a whole-system analysis, which includes the operating system, is generally possible~\cite{schuster:2019:rtas}, we exclude the FreeRTOS components and callbacks into the \system network stack from our analysis for this paper.

We analyze benchmarks including individual driver functions and an evaluation task comprising multiple device interactions.
For each driver function, we determine an upper bound on the execution time.
The evaluation task sends a packet and actively waits for the transmission's completion.
After starting the transmission, the Wi-Fi transceiver transmits the packet and waits for an acknowledgment~(ACK).
Once the transmission is completed, the transmission returns to standby mode. 
We derive an upper bound on both the time and energy demand with respect to the different device states of the Wi-Fi transceiver during the task execution.

As described in Section~\ref{ssec:approach:model}, the static resource analysis requires external knowledge to bound the control flow.
This knowledge includes driver-related and hardware-related facts.
Related to the driver code, we annotate loop bounds based on maximum buffer and packet sizes.
Hardware-related control-flow bounds arise from the driver's need to wait for a hardware event in specific instances, requiring annotations of the waiting loops.
Waiting for the ACK, for example, we annotate with a value of \qty{326}{us}~\footnote{\url{https://sarwiki.informatik.hu-berlin.de/Packet_transmission_time_in_802.11}} for the maximum payload.
Modifying the list of DMA buffers for receiving packets requires waiting for confirmation from the hardware.
We annotate a maximum wait time of \qty{800}{\nano\s}, as observed from our experiments.
While this measurement-based approach may not represent a sound method to obtain worst-case bounds, it demonstrates the feasibility of static resource analysis for our driver.%

The analysis requires a maximum power consumption for the states of the device model shown in Figure~\ref{fig:dev-graph}.
To obtain the power demand, we use the maximum current consumption specified in the ESP32-C3 datasheet~\cite{esp32c3:2022:ds} combined with the voltage of \qty{3.3}{\volt}, which we use for our evaluation setup.
We perform no validation of the datasheet values as the device model can be easily adjusted if the specified values do not reflect the actual worst-case consumption.

In detail, we use the following model:
In state~\sleepstate, the peripheral is switched off.
We assume that this state does not consume any current, i.e., \qty{0}{\milli\ampere}.
In state~\idlestate, the peripheral is switched on.
The device may receive packets, resulting in a current draw of \qty{87}{\milli\ampere}.
In state~\txstate, the device has a current consumption of \qty{285}{\milli\ampere} during transmission.
The baseline energy consumption of the RISC-V CPU with the Wi-Fi peripheral powered down is \qty{28}{\milli\ampere}.

\subsection{Evaluation Results}
\label{sec:eval_results}
\begin{table}[th]
    \centering
    \begin{tabular}{c|>{\raggedleft\arraybackslash}m{1.1cm}|>{\raggedleft\arraybackslash}m{1cm}|>{\raggedleft\arraybackslash}m{1.35cm}}
    \hline
      & \multicolumn{2}{c|}{WCET} & always-on WCEC\\\hline
    \hline
    \texttt{wifi\_hw\_deinit} & 48 cy & 0.3 us & 0.3099 uJ\\
    \texttt{wifi\_setup\_interrupt} & 178 cy & 1.11 us & 1.149 uJ\\
    \texttt{wifi\_setup\_rx} & 1881 cy & 11.8 us & 12.14 uJ\\
    \texttt{wifi\_hw\_init} & 49 cy & 0.306 us & 0.3163 uJ\\
    \texttt{wifi\_transmit\_packet} & 335 cy & 2.09 us & 2.163 uJ\\
    \texttt{wifi\_wait\_for\_tx} & 52184 cy & 326 us & 336.9 uJ\\
    \texttt{wifi\_process\_tx\_done} & 157 cy & 0.981 us & 1.014 uJ\\
    \texttt{wifi\_handle\_rx} & 12989 cy & 81.2 us & 83.85 uJ\\
    \texttt{wifi\_process\_timeout} & 138 cy & 0.862 us & 0.8909 uJ\\
    \texttt{wifi\_get\_bssid} & 94 cy & 0.588 us & 0.6068 uJ\\
    \texttt{wifi\_mac\_handle\_rx} & 68715 cy & 429 us & 443.6 uJ\\
    \texttt{wifi\_interrupt\_handler} & 943 cy & 5.89 us & 6.088 uJ\\\hline
    \end{tabular}
    \caption{Results of the worst-case resource analysis for the individual functions of our open-source Wi-Fi driver.}
    \label{tab:wcec-results}
\end{table}

\begin{table}[th]
    \centering
    \begin{tabular}{c|>{\raggedleft\arraybackslash}m{1.1cm}|>{\raggedleft\arraybackslash}m{1cm}|>{\raggedleft\arraybackslash}m{1.35cm}|>{\raggedleft\arraybackslash}m{1.8cm}}
    \hline
      & \multicolumn{2}{c|}{WCET} & always-on WCEC & device-aware WCEC\\\hline
    \hline
    \texttt{TX task} & 52615 cy & 329 us & 339.7 uJ & 32.42 uJ\\
    \hline
    \end{tabular}
    \caption{Results of the worst-case resource analysis for a task using our open-source Wi-Fi driver for packet transmission.}
    \label{tab:wcec-results-task}
\end{table}

Table~\ref{tab:wcec-results} presents the results of the worst-case resource analysis with Platin for the driver functions and Table~\ref{tab:wcec-results-task} for the transmission task.
The WCET is given in both the number of processor cycles and the physical time (in \si{us}) when operating at a frequency of \qty{160}{\mega\hertz}.
The WCEC is given in \unit{\micro\joule} for two scenarios.
The scenario \texttt{always-on} represents a pessimistic estimate where all devices are assumed to operate in their highest power mode for the entire execution time.
In our evaluations, we consider only the base demand of the processor and the Wi-Fi transceiver as the only additional device.
The scenario \texttt{device-aware}, on the other hand, represents the energy consumption under consideration of the different states of the device during the execution.
As the individual driver functions do not change the device state during their execution, we only give the device-aware WCEC for the task evaluation~(\texttt{TX task}).

Most functions of our open-source Wi-Fi driver perform only operations on memory-mapped registers to control the transceiver and exhibit a relatively low execution time and energy consumption.
The few notably higher resource bounds relate to worst-case analysis assumptions for the control flow.
\texttt{wifi\_mac\_handle\_rx} moves the payload of the received package between buffers, so the analysis has to assume the worst case of the maximum packet size.
Similarly, the analysis has to assume the worst-case number of packets for the execution time of \texttt{wifi\_handle\_rx} and the maximum time until an ACK is received for \texttt{wifi\_wait\_for\_tx}.

The difference between the \texttt{always-on} and \texttt{device-aware} WCEC bound underlines the importance of considering the states of the Wi-Fi transceiver in a fine-granular manner.
We demonstrated that the reverse engineering of our driver code enables deriving the necessary control-flow bounds and performing accurate static resource analysis of Wi-Fi operations on the ESP32-C3.

\section{Related Work}\label{sec:related-work}

\paragraph*{Intermittency-Aware Wireless Communication}
Wireless networking is a key challenge in batteryless designs due to intermittency.
Networking stacks deployed on such devices must be fully aware of intermittent execution~\cite{fu:2023:nobattery}.
In contrast to our work on Wi-Fi, other approaches for batteryless wireless communication use Bluetooth Low Energy (BLE), LoRa, or backscatter-based wireless communications~\cite{cai:2023:bf-wsn,talla:2017:cellphone,jiang:2023:backscatter}.
As a time-predictable IP stack, tpIP~\cite{schoeberl2018tpip} is designed to allow static analysis through bounded loops and avoids dynamic memory allocation.
Similarly, Vogelgesang et al. designed \system, a networking stack with transactional semantics and worst-case bounds for every transaction~\cite{vogelgesang:2025:pfip}.
Transactional semantics break down the overall task into atomic sub-tasks (or transactions) that can be independent of each other.
These approaches are agnostic to the underlying physical layer.

FreeBie~\cite{deWinkel:2022:mobisys} is a bidirectional BLE stack designed for intermittent communication, demonstrated with a batteryless smartwatch.
Cameroptera~\cite{nardello:2019:enssys} is a LoRa-based batteryless image sensing system that adaptively configures its transmission mode based on the available energy.
RockClimb~\cite{choi:2022:rtas} uses WCET bounds without device awareness to guarantee the reaching of checkpoints.
Yarahmadi et al.~\cite{yarahmadi:2020:samos} employ WCEC-aware checkpoint placement without considering the resource demand of peripheral devices.
WoCA introduces device-aware whole-system worst-case analysis for intermittent systems, including a model for LoRa-based communication~\cite{raffeck:2024:lctes}.

Without an intermittency-aware protocol, two devices may not be awake at the same time for communication.
Greentooth~\cite{babatunde:2024:greentooth} synchronizes communication between two devices using time division multiple access (TDMA)-style scheduling and low-power wake-up radios.
Probabilistic modeling of latency or charge times~\cite{geissdorfer:2021:usenix,geissdoerfer:2022:usenix} likewise achieves synchronized communication.

For intermittent devices, Wi-Fi is under-explored due to its high power demand but promises batteryless innovation, for example, in fields like smart-home applications.

\paragraph*{Open-Source Wi-Fi Platforms}
Outside the world of low-power SoCs, open-source drivers exist for several Wi-Fi network interface cards, often supported by the manufacturers themselves, for example, from Atheros or Mediatek~\cite{linux:2025:wireless}.
The OpenWiFi project~\cite{jiao:2020:openwifi} developed an open-source FPGA-based software-defined radio for the 802.11 standard paired with an RF frontend.

On low-power devices, the ZeroWi project~\cite{bentham:2020:zerowi} reverse-engineers the capabilities of the Cypress CYW43438 Wi-Fi SoC, which is programmed using non-free firmware. %
Drivers written for this chip then send commands in the SDIO format over SPI to communicate with the firmware.
Early attempts at reverse engineering Espressif's Wi-Fi driver were performed by Uri Shaked as part of the Wokwi simulator for ESP32~\cite{shaked:2025:wokwi}.
The QEMU-based open-source PICSimLab simulator introduces support for Wi-Fi emulation on the ESP32 and \platform~\cite{gamboa:2025:PICSimlab}.
Based on these attempts, the \openmac project~\cite{devreker:2023:unveiling} aims to fully open-source the Wi-Fi driver on the ESP32.
Our work not only ported \openmac onto the RISC-V--based \platform but also integrated it with
a resource-consumption model for static analysis and
a transactional network stack that is suitable for batteryless devices.

\section{Future Work \& Conclusion}\label{sec:conclusion}

The use of Wi-Fi for intermittently-powered batteryless systems is an ongoing research challenge since network protocols are not designed with intermittent operation in mind~\cite{ahmed:2024:cacm}.
With this work, we contribute to this challenge with our reverse-engineered and publicly-available Wi-Fi drivers, along with their static worst-case energy-consumption analysis.
Our analysis results allow intermittent systems to execute with runtime guarantees on the power-failure--free execution of operations.

Future work needs to consider a more detailed device model to reduce the pessimism in the bounds.
For example, in~\txstate, the peripheral is not actively transmitting after the end of transmission until the ACK is received.
Additionally, energy penalties for switching on/off energy states could be accurately considered.
We also use a static peripheral configuration.
The resource-consumption bounds are impacted by peripheral configurations, such as data rate of transmission.
Future work could identify optimal configurations under energy constraints.
\openmac has recently introduced an initialization of the Wi-Fi peripheral independent of FreeRTOS.
Therefore, an OS-aware implementation could also result in easier static analysis, for example, by exploiting Rust's statically analyzable async/await pattern.

In this work, we presented a fully open-source Wi-Fi driver for the \platform, based on the \openmac project.
We successfully port the project from the Xtensa-based ESP32 to the RISC-V--based \platform using common reverse-engineering techniques.
Finally, we present a proof-of-concept for performing static resource-consumption analysis on the Wi-Fi peripheral's physical operations using the extended Platin worst-case analysis toolkit.

\definecolor{tolerableGreen}{RGB}{40,152,26}
\medskip
\begin{mdframed}[
    backgroundcolor=tolerableGreen!25,
    linecolor=tolerableGreen,linewidth=2pt,topline=false,bottomline=false,
    innertopmargin=0pt, innerbottommargin=3pt, innerleftmargin=6pt, innerrightmargin=6pt,
    frametitleaboveskip=3pt,
    frametitlefont=\itshape, frametitlealignment=\centering,
    frametitle={The implementation is available at: }]
  \centering
  \href{https://git.nt.uni-saarland.de/open-access/pfip}{https://git.nt.uni-saarland.de/open-access/pfip}
\end{mdframed}

\begin{acks}
This work is supported by the German Research Foundation (DFG) as part of SPP 2378 (Resilient Worlds: project number 502615015, \textit{Resilient Power-Constrained Embedded Communication Terminals, ResPECT}) and the NGI0 Core Fund under grant agreement No 101092990.

We would also like to thank all \openmac contributors for their valuable contributions to the community.
\end{acks}

\bibliographystyle{ACM-Reference-Format}
\balance
\bibliography{main.bib}


\begin{thebibliography}{40}

%%% the Web address.

\ifx \showCODEN    \undefined \def \showCODEN     #1{\unskip}     \fi
\ifx \showDOI      \undefined \def \showDOI       #1{#1}\fi
\ifx \showISBNx    \undefined \def \showISBNx     #1{\unskip}     \fi
\ifx \showISBNxiii \undefined \def \showISBNxiii  #1{\unskip}     \fi
\ifx \showISSN     \undefined \def \showISSN      #1{\unskip}     \fi
\ifx \showLCCN     \undefined \def \showLCCN      #1{\unskip}     \fi
\ifx \shownote     \undefined \def \shownote      #1{#1}          \fi
\ifx \showarticletitle \undefined \def \showarticletitle #1{#1}   \fi
\ifx \showURL      \undefined \def \showURL       {\relax}        \fi
% invisible to TeX
\providecommand\bibfield[2]{#2}
\providecommand\bibinfo[2]{#2}
\providecommand\natexlab[1]{#1}
\providecommand\showeprint[2][]{arXiv:#2}

\bibitem[Abella et~al\mbox{.}(2015)]%
        {abella:2015:sies}
\bibfield{author}{\bibinfo{person}{Jaume Abella}, \bibinfo{person}{Carles
  Hern{\'{a}}ndez}, \bibinfo{person}{Eduardo Qui{\~{n}}ones},
  \bibinfo{person}{Francisco~J. Cazorla}, \bibinfo{person}{Philippa~Ryan
  Conmy}, \bibinfo{person}{Mikel Azkarate{-}askasua}, \bibinfo{person}{Jon
  P{\'{e}}rez}, \bibinfo{person}{Enrico Mezzetti}, {and}
  \bibinfo{person}{Tullio Vardanega}.} \bibinfo{year}{2015}\natexlab{}.
\newblock \showarticletitle{{WCET} analysis methods: Pitfalls and challenges on
  their trustworthiness}. In \bibinfo{booktitle}{\emph{Proceedings of the 10th
  International Symposium on Industrial Embedded Systems (SIES '15)}}.
  \bibinfo{pages}{39--48}.
\newblock
\urldef\tempurl%
\url{https://doi.org/10.1109/SIES.2015.7185039}
\showDOI{\tempurl}


\bibitem[Agency(2019)]%
        {nsa:2019:ghidra}
\bibfield{author}{\bibinfo{person}{National~Security Agency}.}
  \bibinfo{year}{2019}\natexlab{}.
\newblock \bibinfo{title}{Ghidra}.
\newblock \bibinfo{howpublished}{https://ghidra-sre.org/}.
\newblock


\bibitem[Ahmed et~al\mbox{.}(2024)]%
        {ahmed:2024:cacm}
\bibfield{author}{\bibinfo{person}{Saad Ahmed}, \bibinfo{person}{Bashima
  Islam}, \bibinfo{person}{Kasim~Sinan Yildirim}, \bibinfo{person}{Marco
  Zimmerling}, \bibinfo{person}{Przemyslaw Pawelczak},
  \bibinfo{person}{Muhammad~Hamad Alizai}, \bibinfo{person}{Brandon Lucia},
  \bibinfo{person}{Luca Mottola}, \bibinfo{person}{Jacob Sorber}, {and}
  \bibinfo{person}{Josiah~D. Hester}.} \bibinfo{year}{2024}\natexlab{}.
\newblock \showarticletitle{The Internet of Batteryless Things}.
\newblock \bibinfo{journal}{\emph{Commun. ACM}} \bibinfo{volume}{67},
  \bibinfo{number}{3} (\bibinfo{year}{2024}), \bibinfo{pages}{64--73}.
\newblock
\urldef\tempurl%
\url{https://doi.org/10.1145/3624718}
\showDOI{\tempurl}


\bibitem[Babatunde et~al\mbox{.}(2024)]%
        {babatunde:2024:greentooth}
\bibfield{author}{\bibinfo{person}{Simeon Babatunde}, \bibinfo{person}{Arwa
  Alsubhi}, \bibinfo{person}{Josiah~D. Hester}, {and} \bibinfo{person}{Jacob
  Sorber}.} \bibinfo{year}{2024}\natexlab{}.
\newblock \showarticletitle{Greentooth: {{Robust}} and {{Energy Efficient
  Wireless Networking}} for {{Batteryless Devices}}}.
\newblock \bibinfo{journal}{\emph{ACM Transactions on Sensor Networks}}
  \bibinfo{volume}{20}, \bibinfo{number}{3} (\bibinfo{year}{2024}),
  \bibinfo{pages}{66:1--66:31}.
\newblock
\urldef\tempurl%
\url{https://doi.org/10.1145/3649221}
\showDOI{\tempurl}


\bibitem[Bentham(2020)]%
        {bentham:2020:zerowi}
\bibfield{author}{\bibinfo{person}{Jeremy Bentham}.}
  \bibinfo{year}{2020}\natexlab{}.
\newblock \bibinfo{title}{Zerowi: Bare-Metal {{WiFi}} Driver for the
  {{Raspberry Pi}}}.
\newblock
\newblock
\urldef\tempurl%
\url{https://iosoft.blog/zerowi/}
\showURL{%
\tempurl}


\bibitem[Berkelaar et~al\mbox{.}(2004)]%
        {lpsolve}
\bibfield{author}{\bibinfo{person}{Michel Berkelaar}, \bibinfo{person}{Kjell
  Eikland}, {and} \bibinfo{person}{Peter Notebaert}.}
  \bibinfo{year}{2004}\natexlab{}.
\newblock \bibinfo{title}{lp\_solve 5.5, open source (mixed-integer) linear
  programming system. Software}.
\newblock
\newblock


\bibitem[Cai et~al\mbox{.}(2023)]%
        {cai:2023:bf-wsn}
\bibfield{author}{\bibinfo{person}{Zhipeng Cai}, \bibinfo{person}{Quan Chen},
  \bibinfo{person}{Tuo Shi}, \bibinfo{person}{Tongxin Zhu},
  \bibinfo{person}{Kunyi Chen}, {and} \bibinfo{person}{Yingshu Li}.}
  \bibinfo{year}{2023}\natexlab{}.
\newblock \showarticletitle{Battery-{{Free Wireless Sensor Networks}}: {{A
  Comprehensive Survey}}}.
\newblock \bibinfo{journal}{\emph{IEEE Internet of Things Journal}}
  \bibinfo{volume}{10} (\bibinfo{year}{2023}), \bibinfo{pages}{5543--5570}.
\newblock
\showISSN{2327-4662}
\urldef\tempurl%
\url{https://doi.org/10.1109/JIOT.2022.3222386}
\showDOI{\tempurl}


\bibitem[Choi et~al\mbox{.}(2022)]%
        {choi:2022:rtas}
\bibfield{author}{\bibinfo{person}{Jongouk Choi}, \bibinfo{person}{Larry
  Kittinger}, \bibinfo{person}{Qingrui Liu}, {and} \bibinfo{person}{Changhee
  Jung}.} \bibinfo{year}{2022}\natexlab{}.
\newblock \showarticletitle{{Compiler-Directed High-Performance Intermittent
  Computation with Power Failure Immunity}}. In
  \bibinfo{booktitle}{\emph{Proceedings of the 28th Real-Time and Embedded
  Technology and Applications Symposium (RTAS '22)}}. \bibinfo{pages}{40--54}.
\newblock
\urldef\tempurl%
\url{https://doi.org/10.1109/RTAS54340.2022.00012}
\showDOI{\tempurl}


\bibitem[de~Winkel et~al\mbox{.}(2022)]%
        {deWinkel:2022:mobisys}
\bibfield{author}{\bibinfo{person}{Jasper de Winkel}, \bibinfo{person}{Haozhe
  Tang}, {and} \bibinfo{person}{Przemyslaw Pawelczak}.}
  \bibinfo{year}{2022}\natexlab{}.
\newblock \showarticletitle{Intermittently-Powered Bluetooth That Works}. In
  \bibinfo{booktitle}{\emph{Proceedings of the 20th Annual International
  Conference on Mobile Systems, Applications and Services (MobiSys '22)}}
  (Portland, Oregon) \emph{(\bibinfo{series}{MobiSys '22})}.
  \bibinfo{pages}{287--301}.
\newblock
\showISBNx{9781450391856}
\urldef\tempurl%
\url{https://doi.org/10.1145/3498361.3538934}
\showDOI{\tempurl}


\bibitem[Dengler et~al\mbox{.}(2023)]%
        {dengler:2023:ecrts}
\bibfield{author}{\bibinfo{person}{Eva Dengler}, \bibinfo{person}{Phillip
  Raffeck}, \bibinfo{person}{Simon Schuster}, {and} \bibinfo{person}{Peter
  Wägemann}.} \bibinfo{year}{2023}\natexlab{}.
\newblock \showarticletitle{{FusionClock}: Energy-Optimal Clock-Tree
  Reconfigurations for Energy-Constrained Real-Time Systems}. In
  \bibinfo{booktitle}{\emph{Proceedings of the 35th Euromicro Conference on
  Real-Time Systems (ECRTS '23)}}, Vol.~\bibinfo{volume}{262}.
  \bibinfo{pages}{6:1--6:24}.
\newblock
\urldef\tempurl%
\url{https://doi.org/10.4230/DARTS.9.1.2}
\showDOI{\tempurl}


\bibitem[Devreker(2023)]%
        {devreker:2023:unveiling}
\bibfield{author}{\bibinfo{person}{Jasper Devreker}.}
  \bibinfo{year}{2023}\natexlab{}.
\newblock \bibinfo{title}{{Unveiling secrets of the ESP32: creating an
  open-source MAC Layer}}.
\newblock
\newblock
\urldef\tempurl%
\url{https://zeus.ugent.be/blog/23-24/open-source-esp32-wifi-mac/}
\showURL{%
Retrieved 2025-01-03 from \tempurl}


\bibitem[{Espressif Systems}(2022)]%
        {esp32c3:2022:ds}
\bibfield{author}{\bibinfo{person}{{Espressif Systems}}.}
  \bibinfo{year}{2022}\natexlab{}.
\newblock \bibinfo{booktitle}{\emph{ESP32-C3 Series Datasheet}}.
\newblock
\urldef\tempurl%
\url{https://www.espressif.com/sites/default/files/documentation/esp32-c3\_datasheet\_en.pdf}
\showURL{%
\tempurl}


\bibitem[{Espressif Systems}(2025a)]%
        {espressif:esp-idf}
\bibfield{author}{\bibinfo{person}{{Espressif Systems}}.}
  \bibinfo{year}{2025}\natexlab{a}.
\newblock \bibinfo{booktitle}{\emph{{ESP IoT Development Framework}}}.
\newblock
\urldef\tempurl%
\url{https://www.espressif.com/en/products/sdks/esp-idf}
\showURL{%
\tempurl}


\bibitem[{Espressif Systems}(2025b)]%
        {esp32c3:trm}
\bibfield{author}{\bibinfo{person}{{Espressif Systems}}.}
  \bibinfo{year}{2025}\natexlab{b}.
\newblock \bibinfo{booktitle}{\emph{ESP32-C3 Technical Reference Manual}}.
\newblock
\urldef\tempurl%
\url{https://www.espressif.com/sites/default/files/documentation/esp32-c3\_technical\_reference\_manual\_en.pdf}
\showURL{%
\tempurl}


\bibitem[Fu et~al\mbox{.}(2023)]%
        {fu:2023:nobattery}
\bibfield{author}{\bibinfo{person}{Shen Fu}, \bibinfo{person}{Vishak
  Narayanan}, \bibinfo{person}{Mathew~L. Wymore}, \bibinfo{person}{Vishal
  Deep}, \bibinfo{person}{Henry Duwe}, {and} \bibinfo{person}{Daji Qiao}.}
  \bibinfo{year}{2023}\natexlab{}.
\newblock \showarticletitle{No battery, no problem: Challenges and
  opportunities in batteryless intermittent networks}.
\newblock \bibinfo{journal}{\emph{Journal of Communications and Networks}}
  \bibinfo{volume}{25}, \bibinfo{number}{6} (\bibinfo{year}{2023}),
  \bibinfo{pages}{806--813}.
\newblock
\urldef\tempurl%
\url{https://doi.org/10.23919/jcn.2023.000033}
\showDOI{\tempurl}


\bibitem[Geissdoerfer and Zimmerling(2021)]%
        {geissdorfer:2021:usenix}
\bibfield{author}{\bibinfo{person}{Kai Geissdoerfer} {and}
  \bibinfo{person}{Marco Zimmerling}.} \bibinfo{year}{2021}\natexlab{}.
\newblock \showarticletitle{Bootstrapping {{Battery-free Wireless Networks}}:
  {{Efficient Neighbor Discovery}} and {{Synchronization}} in the {{Face}} of
  {{Intermittency}}}. In \bibinfo{booktitle}{\emph{18th {{USENIX Symposium}} on
  {{Networked Systems Design}} and {{Implementation}} ({{NSDI}} 21)}}.
  \bibinfo{pages}{439--455}.
\newblock
\showISBNx{978-1-939133-21-2}
\urldef\tempurl%
\url{https://www.usenix.org/conference/nsdi21/presentation/geissdoerfer}
\showURL{%
\tempurl}


\bibitem[Geissdoerfer and Zimmerling(2022)]%
        {geissdoerfer:2022:usenix}
\bibfield{author}{\bibinfo{person}{Kai Geissdoerfer} {and}
  \bibinfo{person}{Marco Zimmerling}.} \bibinfo{year}{2022}\natexlab{}.
\newblock \showarticletitle{Learning to {{Communicate Effectively Between
  Battery-free Devices}}}. In \bibinfo{booktitle}{\emph{19th {{USENIX
  Symposium}} on {{Networked Systems Design}} and {{Implementation}} ({{NSDI}}
  22)}}. \bibinfo{pages}{419--435}.
\newblock
\urldef\tempurl%
\url{https://www.usenix.org/conference/nsdi22/presentation/geissdoerfer}
\showURL{%
\tempurl}


\bibitem[{Gurobi Optimization, LLC}(2025)]%
        {gurobi}
\bibfield{author}{\bibinfo{person}{{Gurobi Optimization, LLC}}.}
  \bibinfo{year}{2025}\natexlab{}.
\newblock \bibinfo{booktitle}{\emph{Reference Manual}}.
\newblock
\urldef\tempurl%
\url{gurobi.com}
\showURL{%
\tempurl}


\bibitem[Jiang et~al\mbox{.}(2023)]%
        {jiang:2023:backscatter}
\bibfield{author}{\bibinfo{person}{Tao Jiang}, \bibinfo{person}{Yu Zhang},
  \bibinfo{person}{Wenyuan Ma}, \bibinfo{person}{Miaoran Peng},
  \bibinfo{person}{Yuxiang Peng}, \bibinfo{person}{Mingjie Feng}, {and}
  \bibinfo{person}{Guanghua Liu}.} \bibinfo{year}{2023}\natexlab{}.
\newblock \showarticletitle{Backscatter Communication Meets Practical
  Battery-Free Internet of Things: {A} Survey and Outlook}.
\newblock \bibinfo{journal}{\emph{IEEE Communications Surveys \& Tutorials}}
  \bibinfo{volume}{25}, \bibinfo{number}{3} (\bibinfo{year}{2023}),
  \bibinfo{pages}{2021--2051}.
\newblock
\urldef\tempurl%
\url{https://doi.org/10.1109/COMST.2023.3278239}
\showDOI{\tempurl}


\bibitem[Jiao et~al\mbox{.}(2020)]%
        {jiao:2020:openwifi}
\bibfield{author}{\bibinfo{person}{Xianjun Jiao}, \bibinfo{person}{Wei Liu},
  \bibinfo{person}{Michael~T. Mehari}, \bibinfo{person}{Muhammad Aslam}, {and}
  \bibinfo{person}{Ingrid Moerman}.} \bibinfo{year}{2020}\natexlab{}.
\newblock \showarticletitle{openwifi: a free and open-source {IEEE802.11} {SDR}
  implementation on SoC}. In \bibinfo{booktitle}{\emph{Proceedings of the 91st
  {IEEE} Vehicular Technology Conference (VTC 2020)}}. IEEE,
  \bibinfo{pages}{1--2}.
\newblock
\urldef\tempurl%
\url{https://doi.org/10.1109/VTC2020-Spring48590.2020.9128614}
\showDOI{\tempurl}


\bibitem[Kim et~al\mbox{.}(2024)]%
        {kim:2024:adma}
\bibfield{author}{\bibinfo{person}{Young-Jun Kim}, \bibinfo{person}{Jiho Lee},
  \bibinfo{person}{Joon-Ha Hwang}, \bibinfo{person}{Youngwook Chung},
  \bibinfo{person}{Byung-Joon Park}, \bibinfo{person}{Junho Kim},
  \bibinfo{person}{So-Hee Kim}, \bibinfo{person}{Junseung Mun},
  \bibinfo{person}{Hong-Joon Yoon}, \bibinfo{person}{Sung-Min Park},
  {et~al\mbox{.}}} \bibinfo{year}{2024}\natexlab{}.
\newblock \showarticletitle{High-Performing and Capacitive-Matched
  Triboelectric Implants Driven by Ultrasound}.
\newblock \bibinfo{journal}{\emph{Advanced Materials}} \bibinfo{volume}{36},
  \bibinfo{number}{2} (\bibinfo{year}{2024}), \bibinfo{pages}{2307194}.
\newblock
\urldef\tempurl%
\url{https://doi.org/10.1002/adma.202307194}
\showDOI{\tempurl}


\bibitem[{lcgamboa}(2022)]%
        {gamboa:2025:PICSimlab}
\bibfield{author}{\bibinfo{person}{{lcgamboa}}.}
  \bibinfo{year}{2022}\natexlab{}.
\newblock \bibinfo{booktitle}{\emph{Picsimlab}}.
\newblock
\urldef\tempurl%
\url{https://lcgamboa.github.io/}
\showURL{%
\tempurl}


\bibitem[{lcgamboa}(2023)]%
        {gamboa:2023:qemu}
\bibfield{author}{\bibinfo{person}{{lcgamboa}}.}
  \bibinfo{year}{2023}\natexlab{}.
\newblock \bibinfo{booktitle}{\emph{Qemu: {{Modified}} Version of the
  {{Espressif QEMU}} Used by the {{PICSimLab}} Simulator.}}
\newblock
\urldef\tempurl%
\url{https://github.com/lcgamboa/qemu/tree/picsimlab-esp32}
\showURL{%
\tempurl}


\bibitem[Maroun et~al\mbox{.}(2024)]%
        {maroun:2024:wcet}
\bibfield{author}{\bibinfo{person}{Emad~Jacob Maroun}, \bibinfo{person}{Eva
  Dengler}, \bibinfo{person}{Christian Dietrich}, \bibinfo{person}{Stefan
  Hepp}, \bibinfo{person}{Henriette Herzog}, \bibinfo{person}{Benedikt Huber},
  \bibinfo{person}{Jens Knoop}, \bibinfo{person}{Daniel Wiltsche{-}Prokesch},
  \bibinfo{person}{Peter~P. Puschner}, \bibinfo{person}{Phillip Raffeck},
  \bibinfo{person}{Martin Schoeberl}, \bibinfo{person}{Simon Schuster}, {and}
  \bibinfo{person}{Peter W{\"{a}}gemann}.} \bibinfo{year}{2024}\natexlab{}.
\newblock \showarticletitle{The Platin Multi-Target Worst-Case Analysis Tool}.
  In \bibinfo{booktitle}{\emph{Proceedings of the 22nd International Workshop
  on Worst-Case Execution Time Analysis (WCET '24)}},
  Vol.~\bibinfo{volume}{121}. \bibinfo{pages}{2:1--2:14}.
\newblock
\urldef\tempurl%
\url{https://doi.org/10.4230/OASIcs.WCET.2024.2}
\showDOI{\tempurl}


\bibitem[Mottola et~al\mbox{.}(2024)]%
        {mottola:2024:eurosec}
\bibfield{author}{\bibinfo{person}{Luca Mottola}, \bibinfo{person}{Arslan
  Hameed}, {and} \bibinfo{person}{Thiemo Voigt}.}
  \bibinfo{year}{2024}\natexlab{}.
\newblock \showarticletitle{Energy Attacks in the Battery-less Internet of
  Things: Directions for the Future}. In \bibinfo{booktitle}{\emph{Proceedings
  of the 17th European Workshop on Systems Security (EuroSec '24)}}.
  \bibinfo{pages}{29--36}.
\newblock
\urldef\tempurl%
\url{https://doi.org/10.1145/3642974.3652283}
\showDOI{\tempurl}


\bibitem[Nardello et~al\mbox{.}(2019)]%
        {nardello:2019:enssys}
\bibfield{author}{\bibinfo{person}{Matteo Nardello}, \bibinfo{person}{Harsh
  Desai}, \bibinfo{person}{Davide Brunelli}, {and} \bibinfo{person}{Brandon
  Lucia}.} \bibinfo{year}{2019}\natexlab{}.
\newblock \showarticletitle{{Camaroptera}: A batteryless long-range remote
  visual sensing system}. In \bibinfo{booktitle}{\emph{Proceedings of the 7th
  International Workshop on Energy Harvesting \& Energy-Neutral Sensing Systems
  (EnsSys '19)}}. \bibinfo{pages}{8--14}.
\newblock
\urldef\tempurl%
\url{https://doi.org/10.1145/3362053.3363491}
\showDOI{\tempurl}


\bibitem[Project(2024)]%
        {linux:2025:wireless}
\bibfield{author}{\bibinfo{person}{Linux Wireless~Documentation Project}.}
  \bibinfo{year}{2024}\natexlab{}.
\newblock \bibinfo{title}{Existing {{Linux Wireless}} Drivers --- {{Linux
  Wireless}} Documentation}.
\newblock
\newblock
\urldef\tempurl%
\url{https://wireless.docs.kernel.org/en/latest/en/users/drivers.html}
\showURL{%
\tempurl}


\bibitem[Raffeck et~al\mbox{.}(2024)]%
        {raffeck:2024:lctes}
\bibfield{author}{\bibinfo{person}{Phillip Raffeck}, \bibinfo{person}{Johannes
  Maier}, {and} \bibinfo{person}{Peter W{\"{a}}gemann}.}
  \bibinfo{year}{2024}\natexlab{}.
\newblock \showarticletitle{{WoCA}: Avoiding Intermittent Execution in Embedded
  Systems by Worst-Case Analyses with Device States}. In
  \bibinfo{booktitle}{\emph{Proceedings of the 25th ACM SIGPLAN/SIGBED
  International Conference on Languages, Compilers, and Tools for Embedded
  Systems (LCTES '24)}}. \bibinfo{pages}{83--94}.
\newblock
\urldef\tempurl%
\url{https://doi.org/10.1145/3652032.3657569}
\showDOI{\tempurl}


\bibitem[Schoeberl(2009)]%
        {schoeberl:2009:eurasip}
\bibfield{author}{\bibinfo{person}{Martin Schoeberl}.}
  \bibinfo{year}{2009}\natexlab{}.
\newblock \showarticletitle{Time-predictable computer architecture}.
\newblock \bibinfo{journal}{\emph{EURASIP Journal on Embedded Systems}}
  \bibinfo{volume}{2009} (\bibinfo{year}{2009}), \bibinfo{pages}{1--17}.
\newblock
\urldef\tempurl%
\url{https://doi.org/10.1155/2009/758480}
\showDOI{\tempurl}


\bibitem[Schoeberl and Pedersen(2018)]%
        {schoeberl2018tpip}
\bibfield{author}{\bibinfo{person}{Martin Schoeberl} {and}
  \bibinfo{person}{Rasmus~Ulslev Pedersen}.} \bibinfo{year}{2018}\natexlab{}.
\newblock \showarticletitle{tpip: A time-predictable tcp/ip stack for
  cyber-physical systems}. In \bibinfo{booktitle}{\emph{2018 IEEE 21st
  International Symposium on Real-Time Distributed Computing (ISORC)}}. IEEE,
  \bibinfo{pages}{75--82}.
\newblock
\urldef\tempurl%
\url{https://doi.org/10.1109/ISORC.2018.00018}
\showDOI{\tempurl}


\bibitem[Schulz(2018)]%
        {schulz:2018:wifi}
\bibfield{author}{\bibinfo{person}{Matthias Schulz}.}
  \bibinfo{year}{2018}\natexlab{}.
\newblock \emph{\bibinfo{title}{Teaching Your Wireless Card New Tricks:
  Smartphone Performance and Security Enhancements Through Wi-Fi Firmware
  Modifications}}.
\newblock \bibinfo{thesistype}{Ph.\,D. Dissertation}.
  \bibinfo{school}{Darmstadt University of Technology, Germany}.
\newblock
\urldef\tempurl%
\url{http://tuprints.ulb.tu-darmstadt.de/7243/}
\showURL{%
\tempurl}


\bibitem[Schuster et~al\mbox{.}(2019)]%
        {schuster:2019:rtas}
\bibfield{author}{\bibinfo{person}{Simon Schuster}, \bibinfo{person}{Peter
  W{\"{a}}gemann}, \bibinfo{person}{Peter Ulbrich}, {and}
  \bibinfo{person}{Wolfgang Schr{\"{o}}der{-}Preikschat}.}
  \bibinfo{year}{2019}\natexlab{}.
\newblock \showarticletitle{{Proving Real-Time Capability of Generic Operating
  Systems by System-Aware Timing Analysis}}. In
  \bibinfo{booktitle}{\emph{Proceedings of the 25th Real-Time and Embedded
  Technology and Applications Symposium (RTAS '19)}}.
  \bibinfo{pages}{318--330}.
\newblock
\urldef\tempurl%
\url{https://doi.org/10.1109/RTAS.2019.00034}
\showDOI{\tempurl}


\bibitem[Shaked(2019)]%
        {shaked:2025:wokwi}
\bibfield{author}{\bibinfo{person}{Uri Shaked}.}
  \bibinfo{year}{2019}\natexlab{}.
\newblock \bibinfo{title}{Wokwi - {{World}}'s Most Advanced {{ESP32
  Simulator}}}.
\newblock
\newblock
\urldef\tempurl%
\url{https://wokwi.com/}
\showURL{%
\tempurl}


\bibitem[Sinha(2024)]%
        {sinha:2024:stateiot}
\bibfield{author}{\bibinfo{person}{Satyajit Sinha}.}
  \bibinfo{year}{2024}\natexlab{}.
\newblock \bibinfo{title}{State of {{IoT}} 2024: {{Number}} of Connected
  {{IoT}} Devices Growing 13\% to 18.8 Billion Globally}.
\newblock
  \bibinfo{howpublished}{https://iot-analytics.com/number-connected-iot-devices/}.
\newblock


\bibitem[Talla et~al\mbox{.}(2017)]%
        {talla:2017:cellphone}
\bibfield{author}{\bibinfo{person}{Vamsi Talla}, \bibinfo{person}{Bryce
  Kellogg}, \bibinfo{person}{Shyamnath Gollakota}, {and}
  \bibinfo{person}{Joshua~R. Smith}.} \bibinfo{year}{2017}\natexlab{}.
\newblock \showarticletitle{Battery-Free Cellphone}.
\newblock \bibinfo{journal}{\emph{Proceedings of the ACM on Interactive,
  Mobile, Wearable and Ubiquitous Technologies}} \bibinfo{volume}{1},
  \bibinfo{number}{2} (\bibinfo{year}{2017}), \bibinfo{pages}{25:1--25:20}.
\newblock
\urldef\tempurl%
\url{https://doi.org/10.1145/3090090}
\showDOI{\tempurl}


\bibitem[Vogelgesang et~al\mbox{.}(2025)]%
        {vogelgesang:2025:pfip}
\bibfield{author}{\bibinfo{person}{Kai Vogelgesang}, \bibinfo{person}{Ishwar
  Mudraje}, \bibinfo{person}{Luis Gerhorst}, \bibinfo{person}{Phillip Raffeck},
  \bibinfo{person}{Peter W{\"a}gemann}, \bibinfo{person}{Thorsten Herfet},
  {and} \bibinfo{person}{Wolfgang {Schr{\"o}der-Preikschat}}.}
  \bibinfo{year}{2025}\natexlab{}.
\newblock \showarticletitle{{{PfIP}}: {{A UDP}}/{{IP Transactional Network
  Stack}} for {{Power-Failure Resilience}} in {{Embedded Systems}}}. In
  \bibinfo{booktitle}{\emph{Proceedings of the Consumer Communications \&
  Networking Conference ({{CCNC}} 2025)}}.
\newblock
\urldef\tempurl%
\url{https://www.nt.uni-saarland.de/publications/vogelgesang2025pfip/PfIP_authors_version.pdf}
\showURL{%
\tempurl}


\bibitem[W\"{a}gemann et~al\mbox{.}(2018)]%
        {waegemann:2018:ecrts}
\bibfield{author}{\bibinfo{person}{Peter W\"{a}gemann},
  \bibinfo{person}{Christian Dietrich}, \bibinfo{person}{Tobias Distler},
  \bibinfo{person}{Peter Ulbrich}, {and} \bibinfo{person}{Wolfgang
  Schr\"{o}der-Preikschat}.} \bibinfo{year}{2018}\natexlab{}.
\newblock \showarticletitle{Whole-System Worst-Case Energy-Consumption Analysis
  for Energy-Constrained Real-Time Systems}. In
  \bibinfo{booktitle}{\emph{Proceedings of the 30th Euromicro Conference on
  Real-Time Systems (ECRTS '18)}}, Vol.~\bibinfo{volume}{106}.
  \bibinfo{pages}{24:1--24:25}.
\newblock
\urldef\tempurl%
\url{https://doi.org/10.4230/LIPIcs.ECRTS.2018.24}
\showDOI{\tempurl}


\bibitem[Wilhelm et~al\mbox{.}(2008)]%
        {wilhelm:2008:tecs}
\bibfield{author}{\bibinfo{person}{Reinhard Wilhelm}, \bibinfo{person}{Jakob
  Engblom}, \bibinfo{person}{Andreas Ermedahl}, \bibinfo{person}{Niklas
  Holsti}, \bibinfo{person}{Stephan Thesing}, \bibinfo{person}{David~B.
  Whalley}, \bibinfo{person}{Guillem Bernat}, \bibinfo{person}{Christian
  Ferdinand}, \bibinfo{person}{Reinhold Heckmann}, \bibinfo{person}{Tulika
  Mitra}, \bibinfo{person}{Frank Mueller}, \bibinfo{person}{Isabelle Puaut},
  \bibinfo{person}{Peter~P. Puschner}, \bibinfo{person}{Jan Staschulat}, {and}
  \bibinfo{person}{Per Stenstr{\"{o}}m}.} \bibinfo{year}{2008}\natexlab{}.
\newblock \showarticletitle{The Worst-case Execution-time Problem -- Overview
  of Methods and Survey of Tools}.
\newblock \bibinfo{journal}{\emph{ACM Transactions on Embedded Computing
  Systems (ACM TECS)}} \bibinfo{volume}{7}, \bibinfo{number}{3}
  (\bibinfo{year}{2008}), \bibinfo{pages}{36:1--36:53}.
\newblock
\urldef\tempurl%
\url{https://doi.org/10.1145/1347375.1347389}
\showDOI{\tempurl}


\bibitem[{Wireshark Foundation}(2025)]%
        {wireshark}
\bibfield{author}{\bibinfo{person}{{Wireshark Foundation}}.}
  \bibinfo{year}{2025}\natexlab{}.
\newblock \bibinfo{title}{free and open-source packet analyzer}.
\newblock \bibinfo{howpublished}{\url{https://www.wireshark.org/}}.
\newblock


\bibitem[Yarahmadi and Rohou(2020)]%
        {yarahmadi:2020:samos}
\bibfield{author}{\bibinfo{person}{Bahram Yarahmadi} {and}
  \bibinfo{person}{Erven Rohou}.} \bibinfo{year}{2020}\natexlab{}.
\newblock \showarticletitle{Compiler optimizations for safe insertion of
  checkpoints in intermittently powered systems}. In
  \bibinfo{booktitle}{\emph{Proceedings of the 20th International Conference on
  Embedded Computer Systems: Architectures, Modeling, and Simulation (SAMOS
  '20)}}, Vol.~\bibinfo{volume}{12471}. \bibinfo{pages}{169--185}.
\newblock
\urldef\tempurl%
\url{https://doi.org/10.1007/978-3-030-60939-9_12}
\showDOI{\tempurl}


\end{thebibliography}

\end{document}